\documentclass{kapproc} 

\usepackage{procps} 
\usepackage{epsf}
\usepackage[dvips]{graphicx}

\kluwerbib 

\begin{document}


\articletitle[Disk Formation]{Gas Rich Mergers in Disk Formation}


\author{Chris~Brook\altaffilmark{1}, Vincent Veilleux\altaffilmark{1}, Daisuke Kawata\altaffilmark{2}, Hugo Martel\altaffilmark{1}, Brad Gibson\altaffilmark{3}}

\altaffiltext{1}{Universit\'{e} Laval $\;^2$ Carnegie Obervatories $\;^3$ Swinburne, ULancashire}

 \begin{abstract}
  In order to explain disk galaxy formation within the hierarchical structure formation, it seems that gas rich mergers must play an important role. We review here our previous studies which have shown the importance of mergers at high redshift  being gas rich, in the formation of both the stellar halo and thick disk components of disk galaxies. Regulation of star formation in the building blocks of our galaxy is required to form a low mass low metallicity stellar halo. This regulation results in high redshift, gas rich mergers during which the thick disk forms. In these proceedings, we categorise stars from our simulated disk galaxy into thin and thick disk components by using the Toomre diagram. Rotation velocity, metallicity and age histograms of the two populations are presented, along with alpha element abundances (oxygen, silicone, magnesium), age-height above the plane, age-radius, metallicity-height, and metalicity-radius gradients.   
\end{abstract}


\begin{section}{Introduction}
The overcooling problem (White \& Rees 1978), coupled with the ineffectiveness of implemented feedback algorithms in regulating star formation, \cite{kg91}, have been central to the difficulties of simulating disk galaxies in a cold dark matter (CDM) Universe. Without sufficient feedback, stars form rapidly in the earliest collapsing dark matter halos, which subsequently merge; angular momentum is transferred to the dark matter halo, and the resultant disk is significantly smaller (in scalelength), and has less angular momentum, than those observed. Feedback from supernovae, stellar winds, quasars etc, has proven to be difficult to simulate, yet the effects of feedback are essential in the formation of disk galaxies. Recent models incorporate these processes without attempting to detail the complex multiphase gas physics involved. This has highlighted the importance of   regulating star formation in the ``building blocks'' of galaxies. (note: We use the term ``building blocks'' when refering to early forming small galaxies which will have merged or been accreted to the final galaxy at redshift 0. By contrast, the term ``satellites'' is reserved for those small galaxies which are found orbiting the central galaxy at z=0). 

To tackle these problems, \cite{tc00} turned off cooling for a fixed time in gas within the Smooth Patricle Hydrodynamical smoothing kernel of a supernova event, resulting in increased angular momentum in the simulated disk galaxy.  Brook et al. (2004a) (see section 2) examined constraints on such feedback models provided by the Milky Way's stellar halo.  Sommer-Larsen and collaborators have examined other implementations of feedback, and also shown the importance of regulating cooling at early epochs e.g. \cite{sl03}. \cite{governato04} highlights the importance of resolution in simulating disk galaxies. Implementing these findings has allowed models such as those presented in this conference by Jesper Sommer-Larsen and Fabio Governato to make progress in reproducing disk galaxies with properties approaching those of observed disks. Although feedback recipes differ, a crucial ingredient in these improved models is the regulation of star formation in galactic building blocks. The implications is that early mergers are gas rich...
\end{section}
    
\begin{section}{Gas Rich Mergers and the Stellar Halo}
In \cite{brooketal04a}, we simulated disk galaxy formation {\it{using identical initial conditions}} with two different feedback methods. Using thermal feedback, which follows \cite{kg91},  star formation is rapid when the building blocks collapse (figure~1), with the stars rapidly reaching relatively high metallicities, and accreting preferentially to the halo. A massive, metal rich stellar halo is formed.   In the adiabatic feedback model, which follows \cite{tc00}, star formation is regulated in low mass building blocks, which remain gas rich when accreted at high redshift. Galaxies form with a low mass, low metallicity stellar halo.  The dissipative gas accretes on to the forming (thick) disk.     

\begin{figure}[h]
\includegraphics[width=7.0cm]{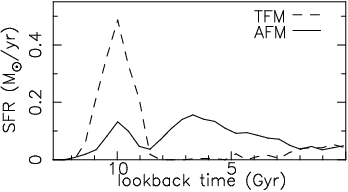}\label{sfrsats}
\narrowcaption{Star formation histories of one galactic building block in the Adiabatic (AFM) and Thermal Feedback Models (TFM). The ability of the AFM to regulate star formation in such building blocks is a crucial ingredient in creating disk galaxies with low mass, low metallicity stellar halos (from Brook et al. 2004a).}
\end{figure}

James Bullock presented a model in which he and his collaborators have modelled the stellar halo using the assumption that stellar halo stars are accreted from  galactic building blocks. They use  cosmologically motivated merger histories to set the timing, mass, and orbits of the  accretions onto an analytic potential. Star formation rates in individal building blocks and satellites are motivated by the mass accretion histories of their host dark matter halos. \cite{bullock} find that star formation in building blocks prior to accretion must be highly regulated, in order to match the properties of the Milky Way's stellar halo; early mergers/accretions are gas rich.  The growth and accretion histories of building blocks are systematically different than those of sattelites, and they show in \cite{font} that such differences  can naturally explain the difference in alpha elements between the stellar halo and the Milky Way's dwarf galaxies (see e.g. the contribution to these proceedings of Kim Venn).
\end{section}

\begin{section}{Gas Rich Mergers and the Thick Disk}
The emerging evidence  that old thick disk stars seem to envelope the majority, and perhaps all disk galaxies, means that the formation of thick disks may be an integral stage in disk galaxy formation. Understanding thick disk formation may thus be a key insight in unravelling disk galaxy formation.      
In  previous papers, we have proposed that the thick disk forms during gas rich mergers, which were frequent at high redshift in a CDM Universe (Brook et al. 2004a, Brook et al. 2005).   This formation scenario was shown to be
consistent with  observations of both the Galactic and extra-galactic
thick disks.  

\begin{figure}[h]
\includegraphics[width=5.0cm,height=4.0cm]{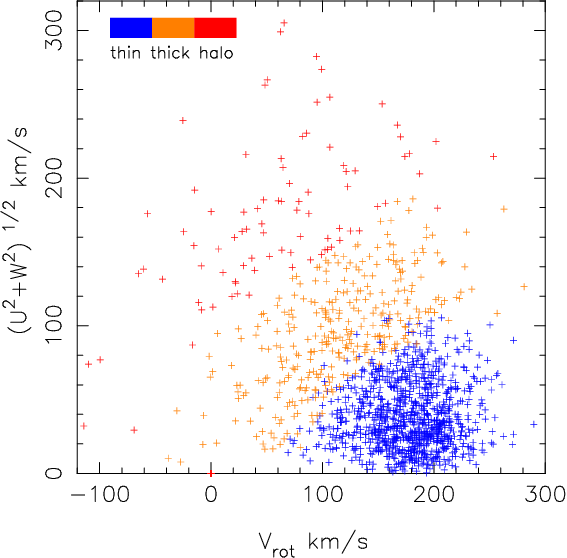}
\narrowcaption{The Toomre diagram of simulation solar neighbourhood stars. Our categoristion of stars as thin disk, thick disk and halo are indicated.}
\end{figure}
 

In these previous studies, we used the abrupt increase in the velocity dispersion-age relation of solar neighbourhood stars, apparent in both the simulations and observations, as an indication of the time of thick disk formation.  By selecting stars according to their age, we were able to relate dynamical and chemical properties of the stellar components, to the dominant processes which were occurring during their birth. This allowed us to show that thick disk stars form during gas rich merger events at high redshift, while disk stars form in a later, more quiescent epoch. In these procedings,  we categorise stars from our simulated disk galaxy into thin and thick disk components by using the Toomre diagram, as shown in figure 2. Rotation velocity, metallicity and age histograms of the two populations are presented (figure~3),  age-height above the plane, age-radius, metallicity-height, and metallicity-radius gradients (figure~4), along with alpha element abundances (oxygen, silicone, magnesium, figure~5). These are consistent with those in \cite{brooketal05} using different selection criteria for the components, in this case better mimicking observational techniques. The conclusions of that paper remain valid.

\begin{figure}[h]
\includegraphics[width=5.0cm,height=4.0cm]{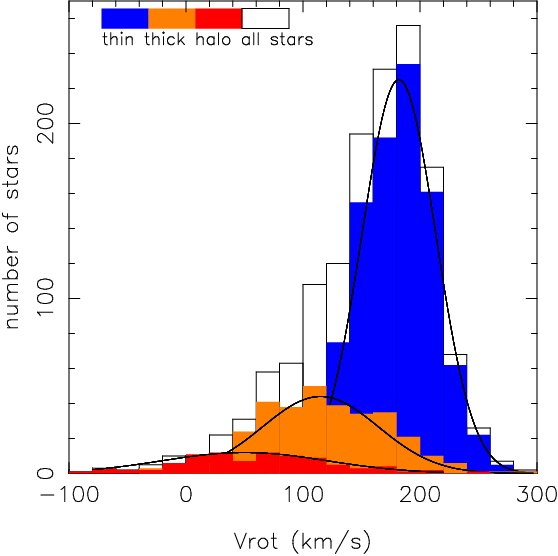}
\hspace{1.0cm}
\includegraphics[width=4.8cm,height=4.cm]{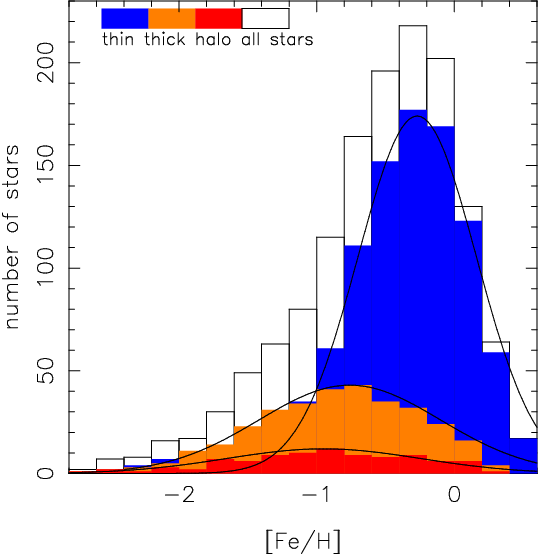}
\end{figure}
\begin{figure}[h]
\includegraphics[width=5.0cm,height=4.0cm]{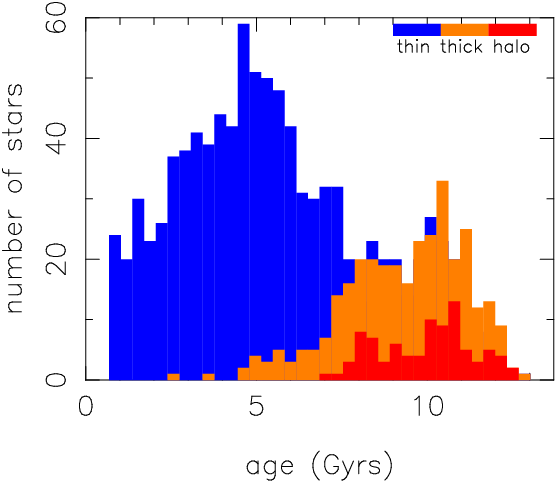}
\narrowcaption{ Rotational velocity ({\it{top left}}), metallicity ({\it{top right}}), and age ({\it{left}}) histograms  of simulation solar neighbourhood stars, depicting disk stars (blue), thick disk (orange) and halo stars (red) as selected from the Toomre diagram.}
\end{figure}

 In these proceedings, Alyson Brooks traces  thick disk stars in their simulated disk galaxy, and finds that only 5-10\% of such stars have been accreted directly from stars in building blocks, implying that direct stellar accretion plays a  minor role in thick disk formation.  Kim Venn  shows that thick disk stars have high $\alpha$ element abundances at relatively high Fe values. This may be explained by the merging epoch, during which the thick disk forms, driving increased star formation rates. The observations of  Debra Elmegreen hint that the progenitors of todays disk galaxies were clumpy, resembling our simulations redshifts at $z\sim 2$. Her interpretation that  Hubble Deep Field disk galaxies are thicker than their local counterparts adds weight to theories such as ours in which thick disks are formed early (Elmegreen et al. 2005).  Julianne Dalcanton presented evidence for the existence of counter-rotating thick disks, (Yoachim \& Dalcanton 2005). This provides further constraints on thick disk formation scenarios. We have a study underway which hopes to determine which thick disk formation scenarios are favoured by such observations.

\begin{figure}
\includegraphics[width=6.5cm]{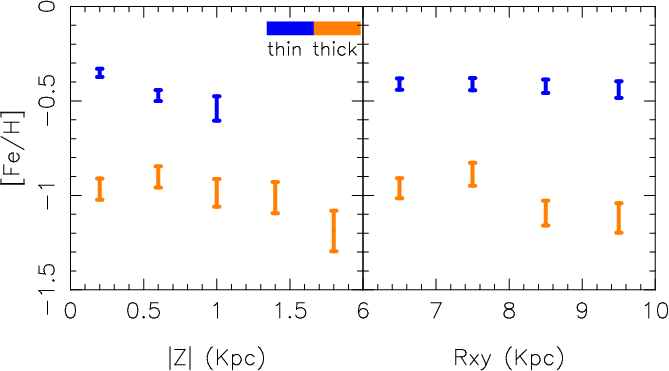}
\includegraphics[width=6.5cm]{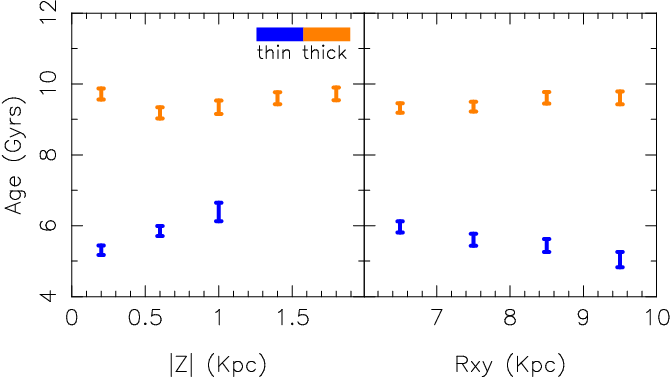}
\caption{ Disk and thick disk metallicity  and age gradients in the radial (R$_{xy}$) and perpendicular (|Z|) directions, showing in particular no gradient with height for thick disk stars.}
\end{figure}

\begin{figure}[h]
\includegraphics[width=7.0cm,height=4.5cm]{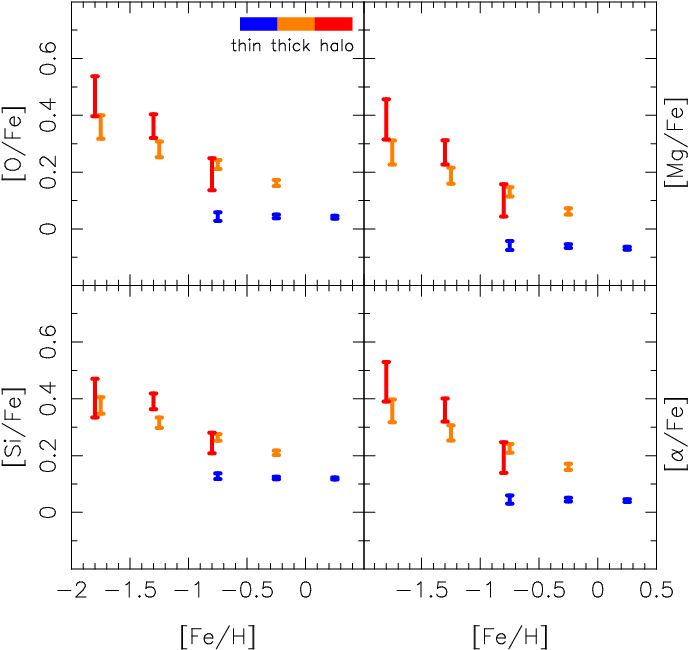}
\narrowcaption{The plots of $\alpha$ elements vs $[Fe/H]$ show that thick disk stars are $\alpha$ enhanced  compared to those in the thin disk.}
\end{figure}

\end{section}  

\begin{section}{Gas Rich Mergers and the Thin Disk}
Feedback at early epochs also results in enrichment of the inter galactic medium, with gas which falls later onto the disk regions being enriched by gas ejected or tidally stripped from  the galactic building blocks, relieving the G-dwarf problem (Brook et al. 2004a). Subsequent ``inside out'' growth of the thin disk in the relatively quiescent period since $z\sim 1$ results in the flattening of the spiral galaxies. This is explored in relation to our simulated disk galaxies in \cite{brooketal06}.

\end{section}
\begin{section}{Conclusion}
If we are to accept $\Lambda$-dominated CDM cosmology, it seems necessary to invoke gas rich mergers to explain many properties of disk galaxies. It is necessary that mergers of building blocks which collapse at high redshift in CDM remain gas rich in creating low mass, low metallicity stellar halos, with observed chemical abundances. The stellar disk that emerges from  the chaotic merging period that characterises CDM between redshifts $\sim 2$ and $\sim 1$, is relatively hot. The rapid star formation induced by these events results in high $\alpha$ abundances in stars where iron abundances are relatively high, in this emergent  thick disk. Gas leftover (gas that is shock heated during this epoch, rather than forming stars) is pre-enriched and falls later onto the thin disk, thus alleviating the G-dwarf problem.  

That gas rich mergers are important in galaxy formation is not a surprise in the sense that it has become clear that, for CDM to work, then the conversion of gas to stars  must become very inefficient as we move toward smaller mass objects.  Such processes are required to explain the multitude of evidence  that is pointing toward ``downsizing'', in which star formation occurs over longer timescales in low mass objects. It must also be remembered that something like 60-70\% of the Milky Way's stars form quiessently from gas in the thin disk, which forms through dissipative processes. We conclude that in low density environments, where disk galaxies form, there is growing evidence that the ``building blocks'' of such galaxies in a CDM Universe remain gas rich. In this manner, there is found a bridge between the two classical galaxy formation paradigms, rapid collapse as proposed in \cite{els}, and the hierarchical build up proposed in \cite{sz} and \cite{wr78}. 
\end{section}


\begin{acknowledgments}
CB thanks  Alyson Brooks, Julianne Dalcanton, Kim Venn, James Bullock,  and Fabio Governato for stimulating discussions and making Island Universes a fun conference. HM and CB are funded through an NSERC grant. 
\end{acknowledgments}

\bibliographystyle{kapalike}
\chapbibliography{<name of .bib file>}


\end{document}